\documentclass[%
 reprint,
 amsmath,amssymb,
 aps,
prb,
]{revtex4-2}

\usepackage{graphicx}
\usepackage{dcolumn}
\usepackage{bm}
\usepackage{braket}
\usepackage{changes}
 \usepackage{color}
\usepackage{comment}
\usepackage{hyperref}       
\hypersetup{
    colorlinks=true,
    linkcolor=blue,
    filecolor=blue,      
    urlcolor=blue,
}
\usepackage{color,soul}

\begin{document}

\preprint{APS/1}

\title{Manipulating multiple optical parametric processes in photonic topological insulators}

\author{Zhen Jiang$^{1,2,\#}$}%
\author{Bo Ji$^{1,2,\#}$}%
\author{Yanghe Chen$^{1,2}$}%
\author{Chun Jiang$^{1}$}%
\email{cjiang@sjtu.edu.cn}	
\author{Guangqiang He$^{1,2}$}%
\email{gqhe@sjtu.edu.cn}	
\affiliation{%
 $^1$State Key Laboratory of Advanced Optical Communication Systems and Networks, Department of Electronic Engineering, Shanghai Jiao Tong University, Shanghai 200240, China\\
 $^2$SJTU Pinghu Institute of Intelligent Optoelectronics, Department of Electronic Engineering, Shanghai Jiao Tong University, Shanghai 200240, China \\
 $^\#$These authors contributed equally to this work.
}%
	
\date{\today}


\maketitle

\textbf{
Topological quantum optics, an emerging area of study, holds the potential to bring about substantial enhancements for integrated quantum devices.
Here we propose integrated topological quantum devices performing various functions including optical parametric amplification, frequency division, and frequency entangled biphoton generation.
We show two distinct edge modes corresponding to different frequency ranges in both sandwich kagome and honeycomb topological designs that emulate the quantum valley Hall effect. 
These two topological edge modes enable two types of optical parametric processes through four-wave mixing, specifically inter-band and intra-band cases.
The devices emulating photonic valley-Hall insulators allow the frequency division of two transverse modes, and furthermore, enable the separation of two quantum functionalities - optical parametric amplification and frequency entangled biphoton state generation.
More importantly, the parametric processes are inborn topological protected, showing robustness against sharp bends and disorders.
Our proposal significantly widens the possibilities for robust, multifunctional topological quantum devices on-chip, which may find applications in quantum information processing.
}

\section{Introduction}
The burgeoning field of on-chip quantum light sources has been undergoing revolutionized development thanks to the advancements in nano-fabrication technologies.
Significant advancements in reducing size and improving stability through photonic integrated circuits have played a pivotal role in enabling the on-chip generation, controlling, and generation of quantum light sources~\cite{1}. 
These improvements have consistently supported more complex and expansive quantum operations, which are crucial for the progress in quantum computing~\cite{2,3}, communication~\cite{4,5,6}, and sensing~\cite{7,8}.
Two key aspects of on-chip quantum light sources are the amplification of light signals and the generation of entangled photon pairs.
Implementing on-chip multifunctional quantum capabilities simultaneously requires precise dispersion engineering and specific materials, which, from a certain perspective, is still challenging.

In parallel, integrating topological phases into quantum systems is enhanced by the robust guidance and manipulation of light, and holds great potential as a cutting-edge and promising area of research~\cite{9,10,11}.
This approach is key to maintaining stable generation and transporting of quantum states.
Topological phases possess a topological nature that grants quantum states with robustness against structural imperfections and disorders. 
Notably, there have been significant advancements in this field, such as the emergence of topological quantum emitters~\cite{12,13}, topological quantum interference~\cite{14,15}, topological biphoton states~\cite{16,17,18}, and even topological quantum frequency combs~\cite{19,20}. 
At the same time, emergent advances in topological nonlinear optics also promise topological protection of complex nonlinear processes~\cite{21,22,23,24,25,26}.
A significant amount of study has been focused on the development of quantum light sources within topological optical systems. 
Nevertheless, the investigation of multifunctional quantum devices in topological photonic systems remains unexplored.

Here we demonstrate integrated topological quantum devices that perform various functions, including optical parametric amplification (OPA), frequency division, and the generation of entangled biphotons. 
We demonstrate the existence of two separate edge modes in sandwich kagome and honeycomb topological designs that emulate the quantum valley Hall (QVH) effect.
By employing a diamond structure, it is possible to couple two separate edge modes to opposite branches, which allows for the separation of spatial modes.
The presence of two edge modes in kagome topological photonic crystals (TPCs) enables four-wave mixing (FWM) processes, leading to two types of optical parametric processes (OPPs) - one related to inter-band scenarios and the other to intra-band scenarios. 
Importantly, the distinct transmission paths of these edge modes enable the individual facilitation of quantum OPA and the generation of continuous frequency entangled biphoton along separate branches.
Besides, we also expand our multifunctional topological quantum devices in honeycomb valley photonic crystals (VPCs).
Additionally, our quantum processes exhibit topological protection features, showing robustness against defects and sharp bends. 
Our approach expands the potential for on-chip, robust, and multifunctional topological quantum devices, opening up new avenues for exploration in quantum optics.

\section{Results} 

\subsection{Topological edge modes in kagome lattice}	
\begin{figure*}
\centering
\includegraphics[width=0.9\textwidth]{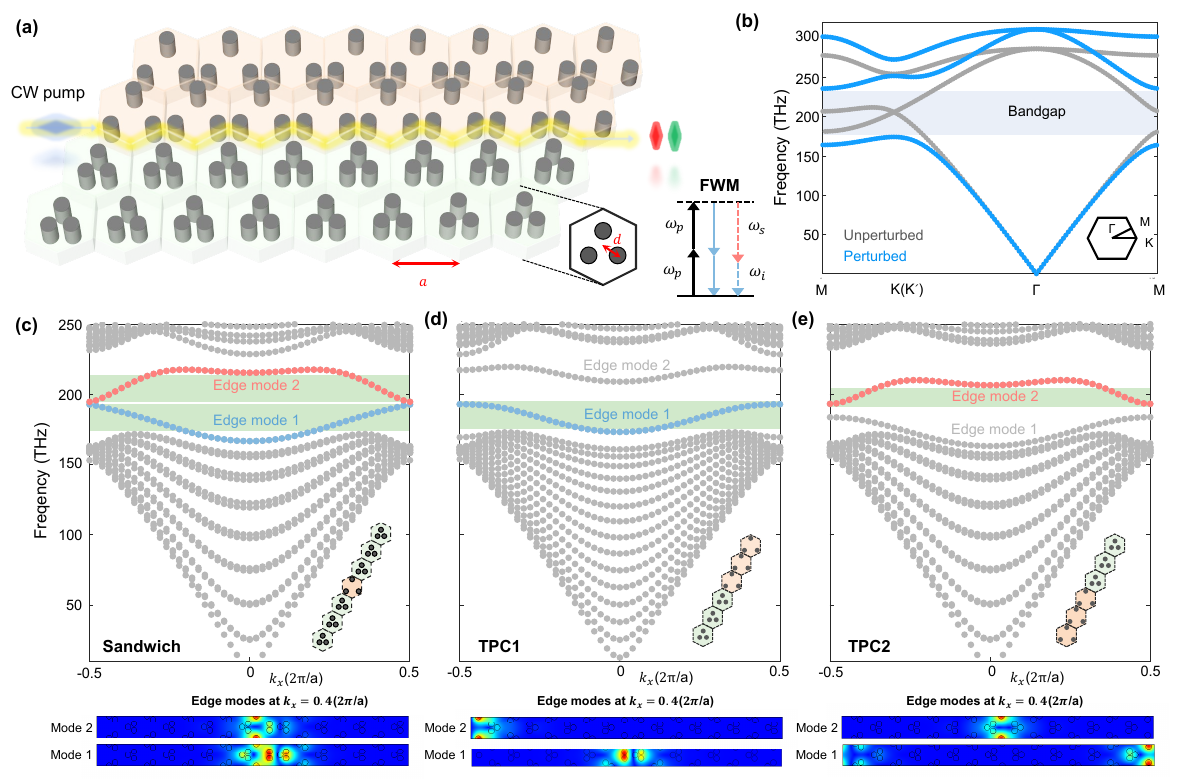}
\caption{(a) A scheme of topological design is composed of two types of silicon-based kagome lattices with $\rm C_3$ symmetry. (b) Band diagrams for an unperturbed kagome lattice (grey dots) and an expanded one (blue dots). Calculated dispersion curves for (c) sandwich TPC, (d) TPC1, and (e) TPC2, respectively, which are composed of shrunken ($d=0.18a$) and expanded ($d=0.40a$) kagome lattices. The bottom insets show the electric field distributions for two edge modes.}
 \label{fig:1}
\end{figure*}

The discovery of the photonic kagome lattice offers a feasible framework for the controllable design of higher-order valley-Hall edge modes~\cite{27,28}. 
Here we explore a two-dimensional topological kagome lattice supporting the generation and flexible control of photonic topological quantum states. 
As depicted in Fig.\ref{fig:1}(a), the topological design consists of two kagome lattices with $\rm C_3$ symmetry.
The effective topological transition is performed by expanding or shrinking an unperturbed kagome lattice with a lattice constant $a = 480$ nm. 
Figure \ref{fig:1}(b) shows the band structures for an unperturbed kagome lattice (grey dots) and an expanded one (blue dots). 
Due to the high symmetry of kagome lattices, there appears a Dirac-like degeneracy at the two high symmetry points ($\rm K$ and $\rm K'$ valleys) of the Brillouin zone~\cite{28}. 
The deformation of the unperturbed kagome lattice leads to a complete photonic bandgap and band inversion mechanism.
The breathing kagome lattice exhibits three mirror symmetries: $M_x$ for the x axis, and $M_{\pm}$ for the two lines obtained by rotating the x-axis by $\pm2\pi/3$~\cite{29}. 
The polarization along the $x_i$ axis represents the expectation value of the position     with $p_i=\frac{1}{S}\int_{\mathrm{BZ}}A_id^2\mathbf{k}$, where $A_{i}=-i\langle\psi|\partial_{k_{i}}|\psi\rangle$ dentes the Berry connection with $x_{i}=x,y$~\cite{29}.
The topological bulk polarization describes the shift in the average position of the Wannier center from the center of the unit cell.
It is noted that the topological bulk difference of shrunken and expanded kagome lattice corresponds to $P=(0,0)$ and $P=(1/3,1/3)$ respectively, which denotes a trivial and nontrivial case, respectively (Supplementary Section I). 

Due to the bulk–boundary correspondence, the nontrivial polarization difference leads to topological edge states localized at boundaries between the shrunken and expanded kagome lattices.
Additionally, for a finite structure, the distorted kagome lattice is expected to show higher-order topological states such as zero-dimensional corner states~\cite{28}.
We first calculate the band structure for a TPC supercell composed of trivial and nontrivial kagome lattices as illustrated in Fig.\ref{fig:1}(c). 
The dispersion curve reveals the presence of two edge modes that are localized in the topological bandgap. 
Furthermore, it leads to an exceptionally wide bandwidth that surpasses 40 THz.  
The bottom inset illustrates that these two modes are exclusively confined to the distinct inner interfaces between two types of kagome lattices.  
This behavior arises from the presence of two boundaries in the sandwich TPC, which correspond to two topological transitions in the case of trivial-nontrivial-trivial topology.

We also perform dispersion calculations for two different TPCs including inverted kagome lattices, which exhibit a single topological interface. 
As depicted in Fig.\ref{fig:1}(d)-(e), both of them exhibit two edge bands located within the topological bandgap.
However, the electric field of two edge modes in each TPC is localized at the outer and inner interfaces, respectively. 
Note that the emergence of outer mode is attributed to the application of periodic boundary conditions on the outer boundaries in the simulation model~\cite{28}.
Correspondingly, exchanging two kagome lattices will lead to the reversion of topological edge states due to the inversion of nontrivial polarization difference~\cite{29}. 
The field distribution and band topologies suggest that the two separate edge modes found in the sandwich TPC are associated with the inner boundary states in the other two TPCs.
The mode-matching behavior simplifies the coupling edge modes from the sandwich TPC to other TPCs, rendering the coupling process more straightforward.
Due to the distinct frequency ranges of the two edge modes, the coupling between different modes achieves a frequency-dependent filtering capability.
In other words, the frequency division characteristic enables the realization of multifunctional on-chip topological photonic devices, potentially finding applications in areas such as optical transmission and light source generation.

\begin{figure*}
\centering
\includegraphics[width=1\textwidth]{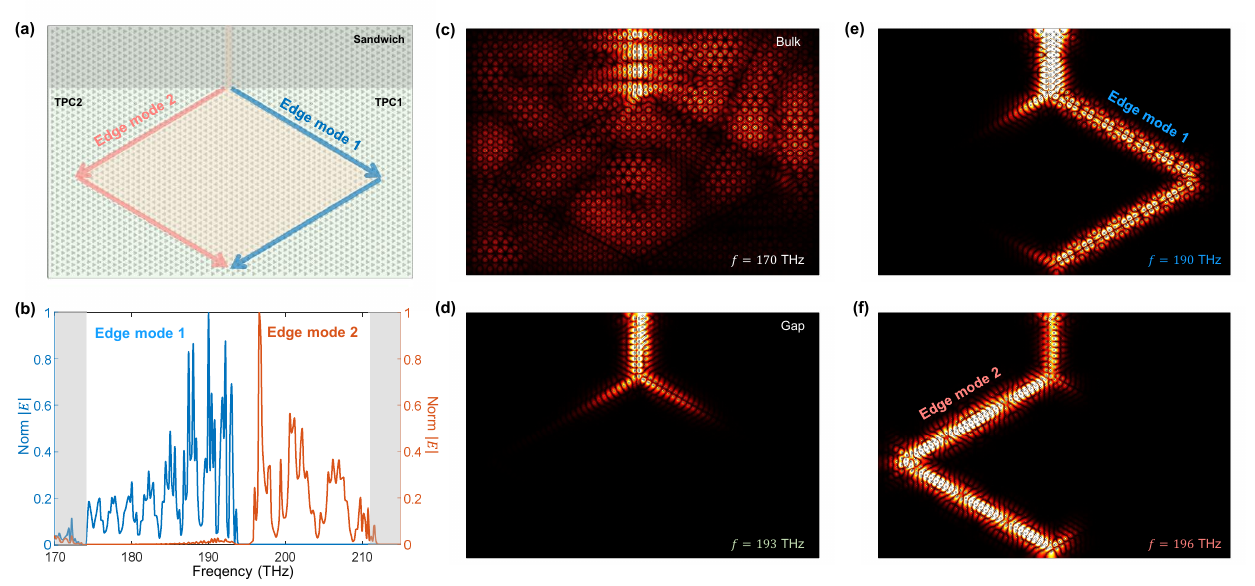}
\caption{(a) A scheme of a topological device composed of a sandwich waveguide (grey region) and a diamond-shaped hybrid structure. 
(b) Normalized field of the light monitored by two probes positioned at the output ports of two branches of the diamond-shaped structure.
(c)-(f) Field profiles for edge modes at different frequencies in our topological device.}
 \label{fig:2}
\end{figure*}

Accordingly, we analyze a sandwich waveguide (depicted as the grey region) alongside a diamond-shaped hybrid structure consisting of trivial (represented by the green region) and nontrivial (represented by the orange region) kagome lattices (Fig.\ref{fig:2}(a)). 
It's important to highlight that in this design, the edges of the rhombus exhibit distinct topological edge states as a result of the mirror symmetry of the lattices. 
There are two allowed edge modes simultaneously with different frequency ranges for the sandwich TPC region. 
However, for the diamond-shaped hybrid structure, the specific structure contains two edge modes characterized by distinct frequency ranges. 
These edge modes can be efficiently transmitted to the left and right branches with the frequency range of $f>193$ THz and $f<193$ THz, respectively.

To get deeper insights into the characteristics of topological edge modes, we simulate the field profiles in this diamond-shaped structure at different frequencies.            
As illustrated in Fig.\ref{fig:2}(c)-(f), the energy couples to the opposite branch corresponding to different pump frequencies. 
Two probes are positioned at the output ports of two branches to monitor field intensity. 
Figure \ref{fig:2}(b) shows the simulated transmission spectra of the light, which clearly reveals the frequency division functionality of our design.
Note that the presence of a frequency gap, where no energy power is detected at either of the output ports, arises from the competition between the two edge modes. 
Correspondingly, as depicted in Fig.\ref{fig:2}(d), the light is unable to interact with any branch at this specific frequency of around 193 THz. 
It is worth mentioning that this dichroic mirror behavior has the potential to facilitate a range of innovative topological functionalities.

\subsection{Multiple OPPs in topological device}
Thanks to the fascinating functionalities of our diamond-shaped topological structure, we expect the stable generation and flexible manipulation of topological quantum states.
As a result of the third-order nonlinearity of silicon, the nonlinear FWM processes generated in sandwich TPCs may lead to the signal and idler photons.
The energy and momentum conversion equations that govern the FWM processes are defined as $2\omega_p=\omega_s+\omega_i$ and $2\boldsymbol{k_p}=\boldsymbol{k_s}+\boldsymbol{k_i}$, where $\omega_{p,s,i}$ and $\boldsymbol{k}_{p,s,i}$ represent the frequencies and wavevectors of the pump, signal, and idler, respectively. Generally, the Hamiltonian for the FWM process in our topological waveguide can be written as 
\begin{equation}
\hat H_{NL}=\hat H_\mathrm{SPM}+\hat H_\mathrm{XPM}+\hat H_\mathrm{FWM},
\label{eq:1}
\end{equation}
where $\hat H_\mathrm{SPM}$, $\hat H_\mathrm{XPM}$ and $\hat H_\mathrm{FWM}$ denote the self-phase modulation (SPM), cross-phase modulation (XPM) and FWM processes, respectively. 
The SPM and XPM terms influence the oscillation process
Due to the frequency division of our diamond-shaped topological structure, the left and right branch corresponds to the OPA process and entangled biphoton generation respectively (see Supplementary Section II for details).

By matching the frequencies utilized in FWM processes with the operating bandwidths of topological edge states, it becomes possible to implement topological protection for entangled biphoton states~\cite{16,17,18}, and even quantum frequency combs~\cite{19,20}.
The dispersion engineering of topological edge states offers a possible method for manipulating FWM processes within the topological bandgap~\cite{16}.
Note that the energy conversion of FWM processes is related to phase-matching intensity $\rm PM=\operatorname{sinc}({\frac{\Delta kL}{2}})$~\cite{32}. 

\begin{figure*}
\centering
\includegraphics[width=1\textwidth]{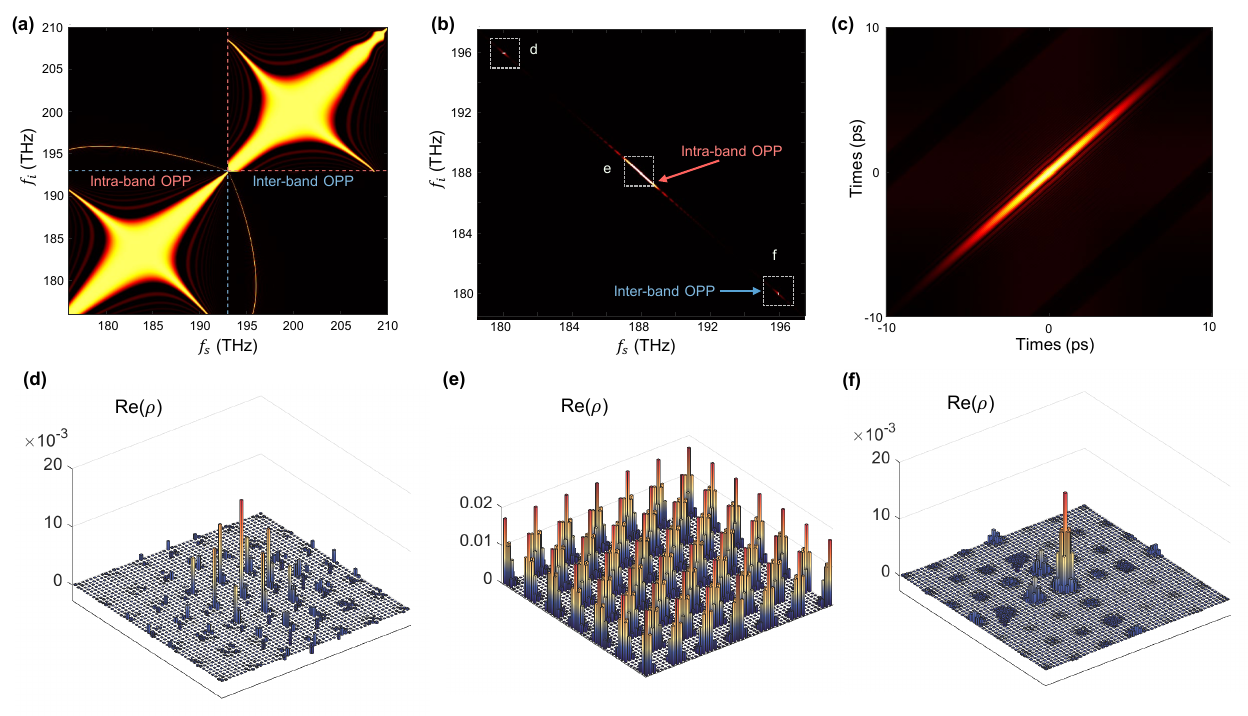}
\caption{(a) Phase-matching intensity distribution of FWM processes for the sandwich TPC. 
(b) JSA distribution and (c) corresponding JTA distribution characterizing the biphoton state generated in the sandwich TPC.
(d)-(f) Three parts of the density matrix $Re(\rho)$ from the JSA of the quantum state.}
\label{fig:3}
\end{figure*}

The phase-matching intensity distribution of FWM processes is depicted in Fig.\ref{fig:3}(a), demonstrating three distinct phase-matching scenarios. 
The two main bright regions correspond to the intra-band OPPs of the two edge modes themselves.
However, besides the intra-band OPP, an additional phase-matching condition is also observed, corresponding to the inter-band OPP between two edge modes.
The nonlinear interactions between two edge modes result in mode conversion, which can lead to significant correlations between different transverse modes.

By pumping the sandwich topological waveguide with the frequency of 188 THz, we can calculate the joint spectral amplitude (JSA) of the biphoton state generated from the FWM process. Such a biphoton state can be given by 
\begin{equation}
|\Psi\rangle=\int\int d\omega_{s}d\omega_{i}{\mathcal A}\left(\omega_{s},\omega_{i}\right)\hat{a}_s^{\dagger}{}{(\omega_{s})}\hat{a}_i^{\dagger}{}{(\omega_{i})}|0\rangle,
\label{eq:1}
\end{equation}
where $\hat{a}^{\dagger}{}_{\omega_{s}}$ and $\hat{a}^{\dagger}{}_{\omega_{s}}$ are creation operators for photons, and ${\mathcal A}\left(\omega_{s},\omega_{i}\right)$ is the JSA. 
The JSA is governed by ${\mathcal A}\left(\omega_{s},\omega_{i}\right)=\alpha(\frac{\omega_{s}+\omega_{i}}{2})\operatorname{sinc}({\frac{\Delta kL}{2}})$, where the pump spectrum $\alpha(\frac{\omega_{s}+\omega_{i}}{2})$ and joint phase-matching spectrum $\operatorname{sinc}({\frac{\Delta kL}{2}})$. 
Our pump is Gaussian with a frequency center of $f_p=188 $ THz and full width at half-maximum of $\Delta f_p=115 $ GHz.

Consequently, the JSA characterizing biphoton state generated in sandwich TPCs is plotted in Fig.\ref{fig:3}(b), where the main intensity region along the diagonal axis denotes a strong signal-idler correlation in the frequency domain~\cite{16}. 
Notably, two extra bright spots are positioned symmetrically above and below the central line, suggesting the presence of inter-band OPP. 
The spot-like phase-matching intensity distribution is also proven to be a perfect case for a heralded single photon generator~\cite{32}, and also, its purity can be improved by machine-learning methods~\cite{33,34}. 
These points indicate the existence of frequency correlations resulting from the additional inter-band OPP interaction between edge mode 1 and edge mode 2.
Note that the signal mode frequency is larger than 193 THz, therefore the generated signal can pass through the left branch while the generated idler passes through the right one.
The potential phase-matching conditions between two different topological edge states promise many effective solutions for manipulating photonic topological quantum states.

As a conjugate variable of frequency, we can obtain the joint temporal amplitude (JTA) of biphotons from the Fourier transformation of JSA by $\tilde{\mathcal A}(t_{s},t_{i})={\mathcal F}[\mathcal A(\omega_{i},\omega_{s})]$~\cite{34}.
The peak intensity of the JTA is located at $\Delta\tau_{y}=\Delta\tau_{z}=0$, indicating a relative phase value of $\phi=0$ (Fig. \ref{fig:3}(c)).
The biphoton state exhibits a signal-idler time correlation with a bandwidth of 10 ps. 
Furthermore, we analyze the photonic topological quantum states by implementing quantum state tomography at a set of bases. 
For the construction of quantum state tomography, the real part of the density matrix is given by $Re(\rho)=\left|\Psi\right\rangle\left<\Psi\right|$, utilizing a basis set comprising 49 frequency divisions. 
We calculate three parts of the density matrix $Re(\rho)$ from the JSA of the quantum state in Fig.\ref{fig:3}(d)-(f) (corresponding to three dashed boxes in Fig.\ref{fig:3}(b)). 

\subsection{Inter-band OPP: tunable OPA}	
\begin{figure*}
\centering
\includegraphics[width=1\textwidth]{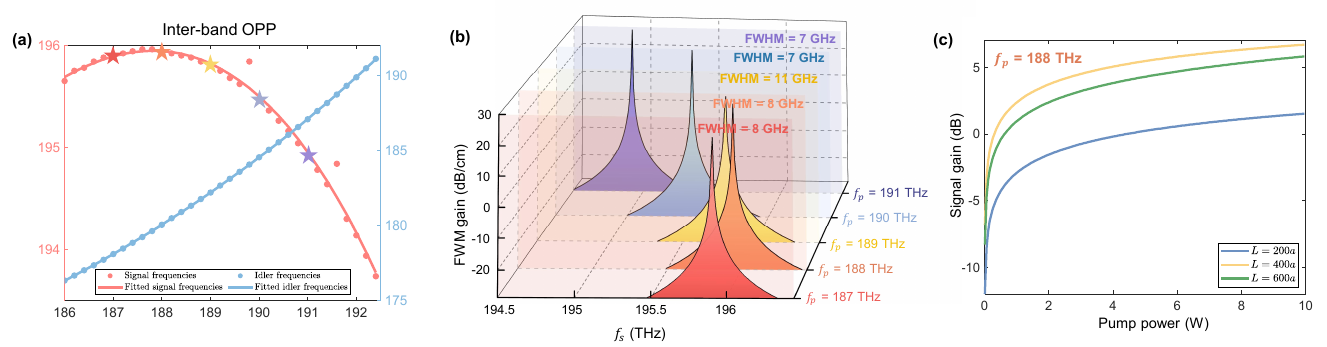}
\caption{(a) Frequency distribution of the signal and idler modes resulting from the inter-band OPP with different pump frequencies.
(b) FWM gain coefficient corresponding to intra-band OPP with varying pump frequencies for a $400a$ length topological waveguide (1 W pump power)
(c) Signal gain as functions of pump power for the waveguide length with $L=200a$, $400a$, and $600a$, respectively.}
 \label{fig:4}
\end{figure*}

Our topological scheme supporting multiple OPPs offers a new approach to manipulating quantum functional devices. 
Here we implement an OPA through FWM with inter-band OPP, the frequency division of the diamond-shaped structure leads to spatial separation of signal photons. 
This spatial separation behavior allows the direct extraction of amplified optical signals since the generated signal could pass through the left branch of the diamond-shaped structure ($f_s>193$ THz).
We initially investigate the frequency distribution of the signal and idler modes resulting from the inter-band OPP with different pump frequencies. 
As shown in Fig.\ref{fig:4}(a), the tunable range of the signal light extends from 193.5 THz to 196 THz, achieving a tunable characteristic of 2.5 THz.

Typically, signal and strong pump modes are coupled into the topological waveguide, in which the signal power is amplified via degenerate FWM~\cite{34-1}.
The FWM gain coefficient is given by $g=\sqrt{\gamma P_{p} \Delta k-\left(\Delta k/2\right)^{2}}$ with effective nonlinearity
$\omega_p{n}_{2}/cA_\mathrm{eff}$, in which $n_2$ is Kerr nonlinearity, $A_\mathrm{eff}$ is the nonlinear effective area.
The effective amplification in the waveguide requires strict adherence to a specific phase-matching condition due to the coherent nature of parametric interaction.
Figure \ref{fig:4}(b) illustrates the FWM gain coefficient corresponding to intra-band OPP with varying pump frequencies for a $400a$ length topological waveguide (1 W pump power).
The intra-band OPP allows a super-narrow bandwidth of large amplification with a half maximum full width (FWHM) of around 8 GHz. 
Also, the device can further achieve tuning of the amplification region by varying the input pump light, ranging from 193.5 THz to 196 THz. 
This tuning process is nonlinear due to the phase-matching condition.
At the central frequency of the amplification region, the FWM gain coefficient of up to 30 dB/cm can be achieved, and the FWM gain is flat during the tuning of the pump frequency.
Such OPA with tunable narrow bandwidth can be applied for amplifying signals from a single photon source.

Consider the pump wave undergoes SPM, simultaneously causing cross-phase modulation XPM on both the signal and idler modes. 
Therefore, the nonlinear phase mismatch caused by SPM and XPM should be taken into account, and the updated phase mismatch is given by $\Delta k_{all}=2\gamma P_{\mathrm{p}}-\Delta k$~\cite{34-2}, where $\gamma=\omega_p n_2A_{eff}$ is the effective nonlinearity of the topological waveguide, $A_{eff}$ is the effective nonlinear interaction area.
Neglecting the optical propagation loss, the observed signal gain generated via FWM for inter-band OPP can be written as~\cite{34-3} 
\begin{equation}
G_{\mathrm{s}}=\frac{P_{\mathrm{s}}^\mathrm{out}}{P_{\mathrm{s}}^\mathrm{in}}=1+\left(\frac{\gamma P_{\mathrm{p}}}{g}\sinh(gL)\right)^2,
\label{eq:2}
\end{equation}
In Fig.\ref{fig:4}(c), we plot the signal gain as functions of pump power for the waveguide length with $L=200a$, $400a$, and $600a$, respectively. 
For a topological waveguide with $L=400a$, a peak gain of 5dB can be achieved when the pump power exceeds 4W.
Note that when the gain of an OPA is large, generated signal photons can undergo significant amplification, reaching macroscopic levels through a phenomenon known as optical parametric generation. The expected number of photons at the output is given by $\langle n\rangle=\sinh^{2}(gL)\approx0.25\exp(2gL)$~\cite{34-4}. 
The detailed quantum analysis of OPA in our topological device is shown in Supplementary Section II. Such a quantum OPA can be used in squeezing light detection~\cite{34-5} and optical homodyne measurement~\cite{34-6}.

\subsection{Inter-band OPP: entangled biphoton state generation}	
\begin{figure}[h]
\centering
\includegraphics[width=0.5\textwidth]{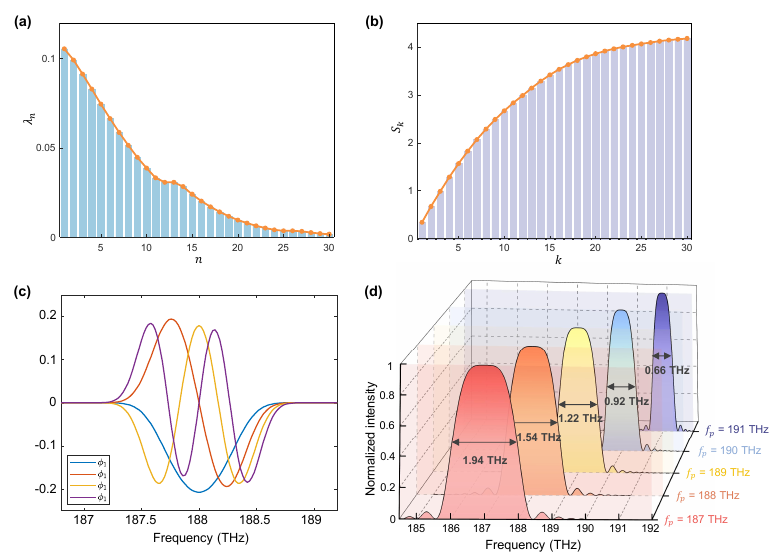}
\caption{Distribution of (a) normalized Schmidt coefficients $\lambda_n$ and (b) entanglement entropy $S_k$ for the biphoton state generated from inter-band OPP.
(c) Eigenfunctions $\phi_n$ ($n = 1,2,3,4$) for the biphoton state.
(d) Normalized two-photon spectral distribution with varying pump frequencies.}
\label{fig:5}
\end{figure}

Furthermore, we can expect the production and manipulation of a frequency entangled biphoton state derived from the inter-band OPP. 
Note that all the pump, signal, and idler modes can couple into the right branch of the diamond-shaped TPC structure ($f_s,f_i<193$ THz), which is convenient for extracting broadband entangled photon pairs directly at this branch.
We employ Schmidt decomposition to evaluate the separability of the JSA without considering part of the phase information~\cite{35,36}. 
Figure \ref{fig:5}(a)-(b) illustrates the distributions of normalized Schmidt coefficients $\lambda_n$ and entanglement entropy $S_k$, respectively.
It is noteworthy that the Schmidt coefficients $\lambda_n$ signify the probability of acquiring the $n$th quantum state, with nonzero coefficients (greater than 1) indicating the frequency entanglement characteristic~\cite{35,37}. 
Moreover, the entanglement entropy, denoted as $S_k = -\sum\lambda_n\log_2\lambda_n$, and the Schmidt number, represented by $K=(\sum\lambda_{n}^{2})^{-1}$, are reliable methods for measuring the degree of entanglement~\cite{36}. 
The entanglement of a topological quantum state can be verified by $S_k > 0$ or $K > 0$, where a higher value of $S_k$ and $K$ indicates a superior quality of frequency entanglement.
For our topological quantum state, the calculated theoretical values for the Schmidt number and entropy of entanglement are $K=16.24$ and $S_{k}=4.42$, respectively, which indicates the emergence of a high-quality frequency entangled biphoton state in our sandwich TPCs.

\begin{figure*}
\centering
\includegraphics[width=0.8\textwidth]{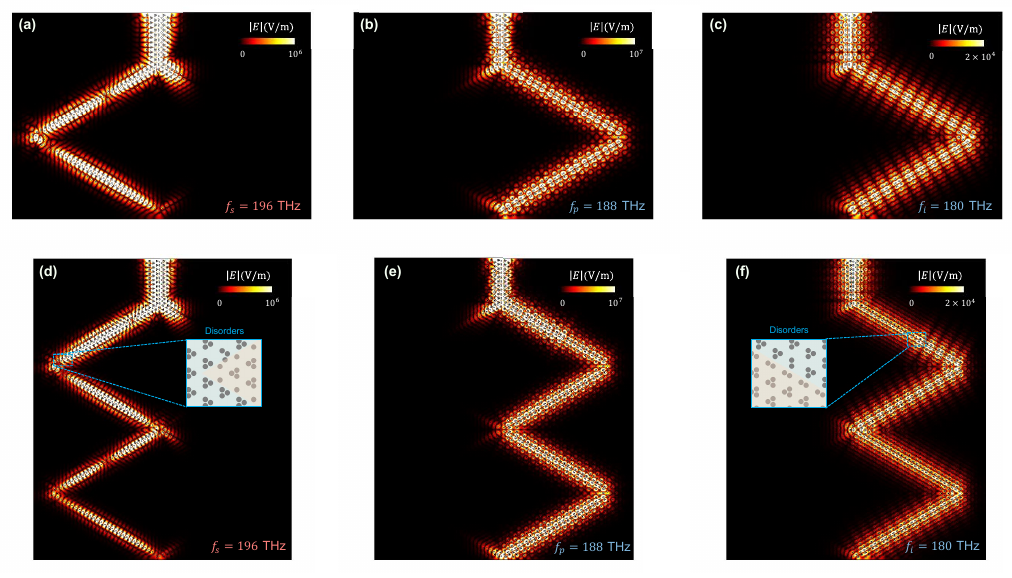}
\caption{Field profiles of the FWM process in the diamond-shaped TPC structure at the frequencies of the (a) signal mode ($f_s=196$ THz), (b) pump mode ($f_p=188$ THz), and (c) idler mode($f_i=180$ THz), respectively. 
Field profiles of the stimulated FWM process in the expanded diamond-shaped TPC structure at the frequencies of the (a) signal mode ($f_s=196$ THz), (b) pump mode ($f_p=188$ THz), and (c) idler mode($f_i=180$ THz), respectively. }
\label{fig:6}
\end{figure*}

Due to the symmetry between the signal and idler photons, the eigenfunctions $\phi$ and $\psi$ in the Schmidt decomposition have the same form. The initial four eigenfunctions $\phi_n$ ($n = 1,2,3,4$) are shown in Fig.\ref{fig:5}(c), which indicates the orthogonality of each basis function.
Also, the number of photon pairs produced from the FWM process is given by
\begin{equation}
\begin{aligned}
S(\omega)& =\langle\Psi|a_{s}^\dagger(\omega_{s})a_{i}^\dagger(\omega_{s})a_{i}(\omega_{s})a_{s}(\omega_{s})|\Psi\rangle\\
&=\frac{\eta^{2}}{c^{2}}\int\mathrm{d}\omega_{\mathrm{s}}\int\mathrm{d}\omega_{\mathrm{i}}|{\mathcal A}(\omega_{\mathrm{s}},\omega_{\mathrm{i}})|^{2},
\label{eq:2}
\end{aligned}
\end{equation}
where $\eta$ is a constant term. Correspondingly, we calculate the normalized two-photon spectral distribution with pump at different frequencies. As shown in Fig.\ref{fig:5}(d), the 3dB bandwidths of the two-photon spectrum are 1.94, 1.54, 1.22, 0.92, and 0.66 THz, respectively, demonstrating the tunability of the spectral bandwidth. 
This high-dimensional topological quantum entangled state with tunable spectral bandwidth enables complex and large-scale quantum simulations and computations.

Alternatively, the single-photon purity associated with the factorizability of biphoton states can be implemented through Schmidt decomposition. 
Purity plays a crucial role in achieving high-visibility quantum interferences between photons emanating from the same source. 
Generally, single-photon purity is expressed as $\text{Tr}(\hat{\rho}_\text{s}^2)$ where $\hat{\rho}_{\mathrm{s}}=\mathrm{Tr}_{\mathrm{i}}(|\Psi\rangle\langle\Psi|)$ represents the density operator for the heralded single photon, and $\mathrm{Tr}{\mathrm{i}}$ is the trace over the idler mode.
The heralded single-photon purity, denoted as $\text{Tr}(\hat{\rho}_\text{s}^2)$, can be calculated by $\text{Tr}(\hat{\rho}_\text{s}^2)\overset{}{=}K^{-1}$~\cite{38},.
Consequently, the single-photon purity for our topological quantum state is computed as 0.06, corresponding to a highly inseparable quantum state.

\subsection{Robustness against disorders for FWM processes}

To verify the topological protection of nonlinear FWM, We simulate the FWM process in the diamond-shaped TPC structure with CW pump excitation~\cite{11, 21} employing COMSOL Multiphysics software.
In our numerical model, we use a point source localized at the input port to excite topological edge modes. 
Notably, there is no input for the idler mode, thereby indicating the excitation of idler modes and the generation of stimulated FWM processes~\cite{11}.
Here the frequencies of the pump, signal, and idler modes were chosen as $f_s=196$ THz, $f_p=188$ THz, and $f_i=180$ THz, respectively.
As depicted in Fig.\ref{fig:6}(a)-(c), the field profiles of topological edge modes at the idler frequency provide clear evidence of the simulated FWM process. 
Most importantly, due to the different frequencies, the pump and signal modes couple into the right branch of the diamond-shaped structure, while the generated idler mode couples into the left branch.
As a result, photon pairs are separated, with an idler photon being extracted during the FWM process.

Additionally, we incorporate a diamond-shaped TPC structure to extend our system's capacity into a larger spatial domain, enabling enhanced manipulation of the transmission routes of photonic topological quantum states.
As depicted in Fig.\ref{fig:6}(d)-(f), within the blue area, a rod's position is randomly moved by distances between -0.1a and 0.2a (left branch), and a rod is randomly removed (right branch).
The idler mode generated via FWM processes also exhibits strong localization along arbitrary topological interfaces. 
These results reveal that the topological nature of the QVH effect brings robustness to the FWM process against sharp bends and defects.
These intriguing behaviors enable the manipulation of the two-photon state's path and the flexible extraction of individual photons.

\section{Discussion}

\subsection{Multiple OPPs in honeycomb lattice}	

\begin{figure*}
\centering
\includegraphics[width=0.9\textwidth]{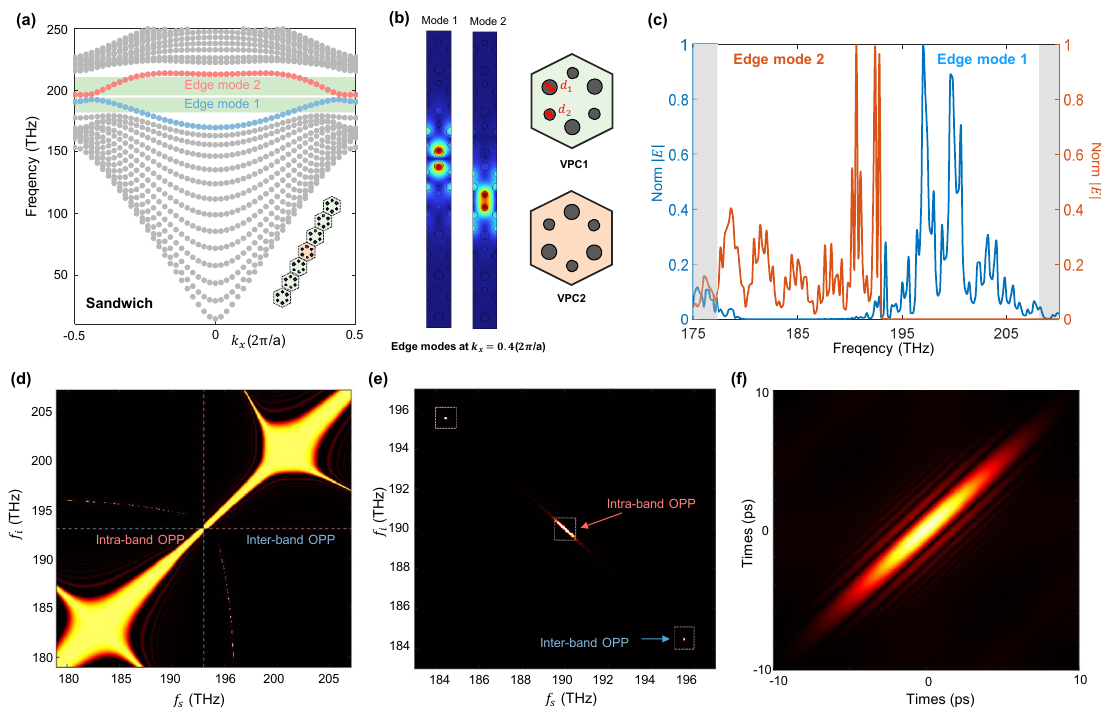}
\caption{Field profiles of the FWM process in the diamond-shaped TPC structure at the frequencies of the (a) signal mode ($f_s=196$ THz), (b) pump mode ($f_p=188$ THz), and (c) idler mode($f_i=188$ THz), respectively. 
Field profiles of the stimulated FWM process in the expanded diamond-shaped TPC structure at the frequencies of the (a) signal mode ($f_s=196$ THz), (b) pump mode ($f_p=188$ THz), and (c) idler mode($f_i=188$ THz), respectively. }
\label{fig:7}
\end{figure*}

We also consider the realization of multifunctional quantum devices in honeycomb VPCs through multiple OPPs.
The valley-Hall honeycomb lattice has been demonstrated as an effective approach to achieving topologically protected edge states~\cite{39,40}, and there is a potential for realizing multiple edge modes.
The VPCs are composed of a honeycomb lattice featuring circular holes with a lattice constant of $a=480$ nm. 
The unperturbed unit cells ($d_1=d_2$) demonstrate a $\rm C_6$ symmetry, resulting in the degeneration of $\rm K$ and $\rm K'$ valleys in the Brillouin zone.
However, breaking the inversion symmetry ($d_1\neq d_2$) leads to a complete bandgap formation near the $\Gamma$ point~\cite{39,40}.
The valley Chern numbers of VPC1 and VPC2 are theoretically calculated as $\rm C_{K/K'}=\pm1/2$~\cite{39,40}, respectively (Supplementary Section III).

Similarly, we calculate the band structure of a sandwich VPC supercell, comprising both nontrivial and trivial honeycomb lattices as shown in Fig.\ref{fig:7}(a). 
The dispersion curve reveals the existence of two edge modes localized within the topological bandgap. 
Figure \ref{fig:7}(b) displays the field distribution for two valley kink states with $k_x=0.4 (2\pi/a)$, revealing that these two modes are localized in two interfaces respectively.
We also simulate the transmission spectra in Fig.\ref{fig:7}(c), which demonstrates the frequency division functionality of our diamond-shaped design.
Thanks to the two edge modes for sandwich VPCs, the phase-matching intensity distribution of FWM processes depicted in Fig.\ref{fig:7}(d) demonstrates two OPPs, including the intra-band OPP and inter-band OPP between two edge modes, respectively.
Correspondingly, the JSA of the biphoton state generated from the FWM process with 190 THz pumping is shown in Fig.\ref{fig:7}(e).
Likewise, there also exist a main region and two symmetrical bright spots, corresponding to intra-band and inter-band OPP, respectively.
These two OPPs are expected to be performed as entangled biphoton generation and an OPA process (Supplementary Section III).
Further, we can obtain the JTA of biphotons from the Fourier transformation of JSA as illustrated in Fig.\ref{fig:7}(f), leading to a signal-idler correlation with a 10 ps bandwidth.
More importantly, we have shown that our topological devices that conduct multiple OPPs can be implemented in valley-Hall kagome and honeycomb lattices.
Therefore, we believe that multifunctional quantum devices can be fabricated in a broader range of topological structures in the future.

\section{Conclusion}
In this work, we demonstrate on-chip topological quantum optical devices capable of performing multiple functions including OPA, frequency separation, and entangled biphoton generation.
We show that there exist two distinct edge modes corresponding to different frequency ranges in sandwich TPCs.
By employing a diamond structure, we can couple these two edge modes into separate branches to achieve the separation of spatial modes.
Due to the coexistence of two edge modes, the FWM process enables two types of OPPs, corresponding to inter-band and intra-band cases, respectively. 
More importantly, thanks to distinct transmission paths of the edge modes, these two OPPs can individually facilitate quantum OPA and the generation of continuous frequency-entangled photon pairs along separate branches. 
Moreover, these quantum processes exhibit topological protection features, demonstrating robustness against defects and sharp bends.
Our proposal offers increased possibilities for on-chip robust, multifunctional topological quantum devices.


\begin{acknowledgments}
This work is supported by the Key-Area Research and Development Program of Guangdong Province (2018B030325002), the National Natural Science Foundation of China (62075129, 61975119), the SJTU Pinghu Institute of Intelligent Optoelectronics (2022SPIOE204), and the Sichuan Provincial Key Laboratory of Microwave Photonics (2023-04).
\end{acknowledgments}

\textbf{Competing interests}
The authors declare no competing interests.


%


\end{document}


\renewcommand{\thefigure}{S\arabic{figure}}
\renewcommand{\theequation}{S\arabic{equation}}
\renewcommand{\thetable}{S\arabic{table}}

\preprint{APS}

\title{Supplementary Materials: Manipulating multiple optical parametric processes in photonic topological insulators}

\author{Zhen Jiang$^{1,2,\#}$}%
\author{Bo Ji$^{1,2,\#}$}%
\author{Yanghe Chen$^{1,2}$}%
\author{Chun Jiang$^{1}$}%
\email{cjiang@sjtu.edu.cn}	
\author{Guangqiang He$^{1,2}$}%
\email{gqhe@sjtu.edu.cn}	
\affiliation{%
 $^1$State Key Laboratory of Advanced Optical Communication Systems and Networks, Department of Electronic Engineering, Shanghai Jiao Tong University, Shanghai 200240, China\\
 $^2$SJTU Pinghu Institute of Intelligent Optoelectronics, Department of Electronic Engineering, Shanghai Jiao Tong University, Shanghai 200240, China \\
 $^\#$These authors contributed equally to this work.
}%

\maketitle

\section{Topological kagome lattice}
\begin{figure*}
\centering
\includegraphics[width=0.7\textwidth]{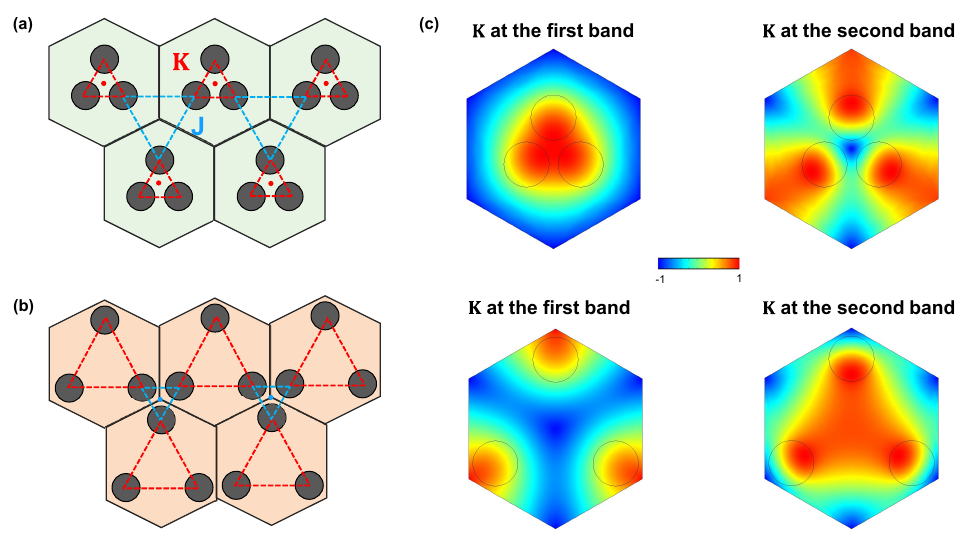}
\caption{(a) Shrunken and (b) expanded kagome lattice for trivial and nontrivial cases. (c) Electric field distributions at point $\rm K$ in the first Brillouin zone for two kagome lattices.}
\label{fig:S1}
\end{figure*}

We consider a two-dimensional infinite kagome lattice that exhibits $\rm C_3$ lattice symmetry (lattice constant $a$), as shown in Fig.\ref{fig:S1}. 
We apply the Tight-Binding model to the lattice, considering only the nearest hopping term in the model. 
As a result, the Hamiltonian model in the momentum space can be given by~\cite{s1}

\begin{equation}
{{\hat H}}_0=\begin{pmatrix}0&K+je^{i(\frac{1}{2}k_x+\frac{\sqrt{3}}{2}k_y)a}&K+je^{-i(\frac{1}{2}k_x-\frac{\sqrt{3}}{2}k_y)a}\\K+je^{-i(\frac{1}{2}k_x+\frac{\sqrt{3}}{2}k_y)a}&0&K+je^{-ik_xa}\\K+je^{i(\frac{1}{2}k_x-\frac{\sqrt{3}}{2}k_y)a}&K+je^{ik_xa}&0\end{pmatrix},
\label{eq:S1}
\end{equation}	
where $K$ and $J$ denote the intra-cell coupling (red dotted line) and inter-cell coupling (blue dotted line), respectively, as depicted in Fig.\ref{fig:S1}(a).
We can write this Hamiltonian in a more general form:
\begin{equation}
{{\hat H}}_0=\begin{pmatrix}0&a_1&b_1\\a_2&0&c_1\\b_2&c_2&0\end{pmatrix},
\label{eq:S2}
\end{equation}	
Next, the generalized chiral symmetry operator in kagome lattices can be described as~\cite{s1}:
\begin{equation}
\Gamma_3=\begin{pmatrix}1&0&0\\0&e^{i2\pi/3}&0\\0&0&e^{-i2\pi/3}\end{pmatrix},
\label{eq:S3}
\end{equation}	
where $\Gamma_3$ is the unitary chiral operator with three eigenvalues of 0, $e^{i2\pi/3}$, and $e^{-i2\pi/3}$.
For the original Hamiltonian $H_0$, the transformed Hamiltonians by the unitary chiral operator are given by~\cite{s2} 
\begin{equation}
{\hat H}_1=\Gamma_3\ H_0\Gamma_3^{-1}=\begin{pmatrix}0&e^{-i2\pi/3}a_1&e^{i2\pi/3}b_1\\e^{i2\pi/3}a_2&0&e^{-i2\pi/3}c_1\\e^{-i2\pi/3}b_2&e^{i2\pi/3}c_2&0\end{pmatrix},
\label{eq:S4}
\end{equation}	
\begin{equation}
{\hat H}_{2}=\Gamma_{3}H_{1}\Gamma_{3}^{-1}=\begin{pmatrix}0&e^{i2\pi/3}a_{1}&e^{-i2\pi/3}b_{1}\\e^{-i2\pi/3}a_{2}&0&e^{i2\pi/3}c_{1}\\e^{i2\pi/3}b_{2}&e^{-i2\pi/3}c_{2}&0\end{pmatrix}.
\label{eq:S5}
\end{equation}	
Therefore, these Hamiltonians satisfy $\hat H_0+\hat H_1+\hat H_2=0$, which reveals that the kagome lattice has generalized chiral symmetry. The generalized chiral symmetry promises that the sum of the respective eigenenergies is equal to zero. 
The current Hamiltonian is analogous to the Hamiltonian in the Su-Schrieffer-Heager (SSH) model.
The introduction of long-range interactions, despite disrupting the chiral symmetry, will lead to a change of topological invariant (bulk polarization).

To describe the band topology of the kagome lattice with complete symmetry, the bulk polarization is defined as~\cite{s3} 
\begin{equation}
p_{s}=1/N_{k}\sum_{j,k_{t}}v_{s}^{j}(k_{t}),
\label{eq:S6}
\end{equation}	
where $v_{s}^{j}(k_{t})$ is the eigenvalue (also denoted as Wannier center) of the Wilson loop. And the eigenvalue problem of the Wilson loops is $W_{k_{s}+2\pi\leftarrow k_{s},k_{t}}|\nu_{k}\rangle^{j}=e^{i2\pi\nu_{s}^{j}(k_{t})}|\nu_{k}\rangle^{j}$, in which $k_{s},k_{t}=0,\delta k,\ldots,(N_{k}-1)\delta k,\delta k=\frac{1}{N_{k}}\frac{4\pi}{\sqrt{3}a}$ and $j$ is the index of the occupied bands.
Hence, the Wannier bands associated with the lowest energy band of kagome lattices can be calculated~\cite{s2}. 
When the intra-cell coupling $K$ is larger than inter-cell coupling $J$ (shrunken lattice), the bulk polarization is calculated as 0, which denotes a trivial case (Fig.\ref{fig:S1}(a)). 
However, when the intra-cell coupling $K$ is smaller than inter-cell coupling $J$ (expanded lattice), the calculated bulk polarization is $1/3$, which denotes a nontrivial case (Fig.\ref{fig:S1}(b)).
Therefore, according to the bulk–boundary correspondence, the difference in nontrivial polarization gives rise to topological edge states that localize at the boundaries between the contracted and expanded kagome lattices.
As shown in Fig.\ref{fig:S1}(c), the electric field distributions at point $K$ in the first Brillouin zone for two kagome lattices also reveal the topological transition.

\section{Theoretical analysis of two OPPs in topological device}
For the FWM process in our topological device, it contains self-phase modulation (SPM), cross-phase modulation (XPM), and FWM processes. 
The general nonlinear Hamiltonian can be written by $\hat H_{NL}=\hat H_\mathrm{SPM}+\hat H_\mathrm{XPM}+\hat H_\mathrm{FWM}$, we can write all the terms as 
\begin{equation}
\begin{aligned}
\hat H_{NL} 
= -\hbar \gamma[& \frac12(\hat{a}_p^{\dagger}\hat{a}_p^{\dagger}\hat{a}_p\hat{a}_p+\hat{a}_s^{\dagger}\hat{a}_s^{\dagger}\hat{a}_s\hat{a}_s+\hat{a}_i^{\dagger}\hat{a}_i^{\dagger}\hat{a}_i\hat{a}_i)\\
 &+2(\hat{a}_p^{\dagger}\hat{a}_s^{\dagger}\hat{a}_p\hat{a}_s+\hat{a}_p^{\dagger}\hat{a}_i^{\dagger}\hat{a}_p\hat{a}_i+\hat{a}_s^{\dagger}\hat{a}_i^{\dagger}\hat{a}_s\hat{a}_i)\\
 &+(\hat{a}_s^{\dagger}\hat{a}_i^{\dagger}\hat{a}_p\hat{a}_p+\hat{a}_p^{\dagger}\hat{a}_p^{\dagger}\hat{a}_s\hat{a}_i)]
\end{aligned}
\label{eq:s7}
\end{equation}	
where $\hat a_{p,s,i}$ and $\hat a_{p,s,i}^\dagger$ and $a_{s,\omega_i}^\dagger$ are the annihilation operator and creation operators respectively, the effective nonlinearity is $\gamma=\omega_p{n}_{2}/cA_\mathrm{eff}$, and $n_2$ is the Kerr nonlinearity of silicon with a value of $5\times 10^{-18}$~\cite{s4}, $A_{eff}$ is the nonlinear effective area.
The final term (FWM process) is crucial for the energy transfer among the three modes. The SPM and XPM terms significantly influence the oscillation process, as well as the noise and entanglement characteristics of the system. Note that we only consider the FWM process generated in sandwich topological waveguides with a length $L=400a$. However, the FWM processes in two branches of the diamond-shaped structure are neglected since the nonlinear interaction length is small.

\subsection{Left branch: OPA process generated from inter-band OPP}
In our setup, we employ optical parametric amplification (OPA) through FWM, combined with inter-band optical parametric processes (OPP). 
This approach, utilized within the diamond-shaped structure, results in the spatial separation of signal photons to the left branch.
For the OPA process, we analyze the evolution of quantum states by solving the Heisenberg equation $\frac{d\hat{a}_{j}}{dt}=\frac{i}{\hbar}[\hat{H},\hat{a}_{j}],j\in\{p,s,i\}$. 
By substituting Eq.\ref{eq:s7}, we can calculate that the updated Heisenberg equations for the signal and idler modes are 
\begin{equation}
\begin{aligned}
\frac{d \hat a_{p}}{dt} &=i\gamma[(\hat{a}_p^{\dagger}\hat{a}_p+2(\hat{a}_s^{\dagger}\hat{a}_s+\hat{a}_i^{\dagger}\hat{a}_i)\hat{a}_p+\hat{a}_s\hat{a}_i\hat{a}_p^{\dagger}]\\ 
\frac{d \hat a_{s}}{dt} &=i\gamma[(\hat{a}_s^{\dagger}\hat{a}_s+2(\hat{a}_p^{\dagger}\hat{a}_p+\hat{a}_i^{\dagger}\hat{a}_i)\hat{a}_s+\hat{a}_i^{\dagger}\hat{a}_p\hat{a}_p]\\ 
\frac{d \hat a_{i}}{dt} &=i\gamma[(\hat{a}_i^{\dagger}\hat{a}_i+2(\hat{a}_p^{\dagger}\hat{a}_p+\hat{a}_s^{\dagger}\hat{a}_s)\hat{a}_i+\hat{a}_s^{\dagger}\hat{a}_p\hat{a}_p]
\end{aligned}
\label{eq:s8}
\end{equation}	
We may explore the interaction of three stationary, co-polarized waves at regular frequencies, characterized by their slowly varying electric fields with complex amplitudes $A_p(x)$, $A_s(x)$, and $A_i(x)$, respectively.
The total transverse field $E(x,y,z)$ propagating along the sandwich topological waveguide (x-axis) is given by~\cite{s5}
\begin{equation}
\begin{aligned}
E(x,y,z) =&f(y,z)A(x)\\
=&f(y,z)\frac{1}{2}[A_{p}(x)  
\times\exp(i k_{0}x-i\omega_{0}t)+A_{s}(x)\exp(i k_{1}x-i\omega_{1}t) \\
&+A_{i}(x)\exp(i k_{2}x-i\omega_{2}t)+h.c.],
\end{aligned}
\label{eq:S9}
\end{equation}	
in which h.c. refers to the complex conjugate. The $f(y,z)$  denotes a common transverse modal profile, which is assumed to be identical for all three waves propagating along the waveguide. 
According to Eq.\ref{eq:s8}, we can derive three coupled equations for the classicized field amplitudes of the three waves as~\cite{s6}
\begin{equation}
\begin{gathered}
\frac{dA_{p}}{dx} =i\gamma[(|A_{p}|^{2}+2(|A_{s}|^{2}+|A_{i}|^{2}))A_{p} 
+2A_{s}A_{i}A_{p}^{*}\exp(i\Delta k x)], \\
\frac{dA_{s}}{dx} =i\gamma[(|A_{s}|^2+2(|A_{i}|^2+|A_{p}|^2))A_{s} 
+A_{i}^{*}A_{p}^{2}\exp(-i\Delta k x)], \\
\frac{dA_i}{dx} =i\gamma[(|A_{i}|^2+2(|A_{s}|^2+|A_{p}|^2))A_{i} 
+A_{s}^{*}A_{p}^{2}\exp(-i\Delta k x)]. 
\end{gathered}
\label{eq:s10}
\end{equation}	
Note that the first two terms on the right-hand side of Eq.\ref{eq:s10} denote the nonlinear phase shifts due to SPM and XPM, respectively. 
The last term denotes energy transfer between the interacting waves.
By replacing the  amplitude of the light field by $A_{j}(x)=\sqrt{P_{j}}\exp(i\phi_{j})\mathrm{~for~}j\in\{p,s,i\}$, Eq.\ref{eq:s10} can be rewritten as~\cite{s6} 
\begin{equation}
\begin{aligned}
\frac{dP_{p}}{dx}& =-4\gamma\left(P_{p}^{2}P_{s}P_{i}\right)^{1/2}\sin\theta,   \\
\frac{dP_{s}}{dx}& =2\gamma\left(P_{p}^{2}P_{s}P_{i}\right)^{1/2}\sin\theta,   \\
\frac{dP_i}{dx}& =2\gamma\left(P_{p}^{2}P_{s}P_{i}\right)^{1/2}\sin\theta,   
\end{aligned}
\label{eq:s11}
\end{equation}	
and 
\begin{equation}
\frac{d\theta}{dx} =\Delta k+\gamma(2P_{p}-P_{s}-P_{i})+\gamma\left[\left(P_{p}^{2}P_{i}/P_{s}\right)^{1/2}\right.  
\left.+\left(P_{p}^{2}P_{i}/P_{s}\right)^{1/2}-4\left(P_{s}P_{i}\right)^{1/2}\right]\cos\theta.
\label{eq:s12}
\end{equation}	
Neglecting the third term in Eq.\ref{eq:s12}, an approximated result for relative phase difference is given by~\cite{s5}
\begin{equation}
\frac{d\theta}{dx}\approx\Delta k+\gamma(2P_{p}-P_{s}-P_{i})\approx\Delta k+2\gamma P_{p}.
\label{eq:s13}
\end{equation}	

Eq.\ref{eq:s10} describes the amplification of a weak signal propagating along the topological waveguide. To solve the equations, let $\frac{d A_0}{dx}=0$, then we can get an analytical solution~\cite{s7}
\begin{equation}
P_{s}(L) =P_s(0)\left(1+\left[\frac{\gamma P_p}{g}\sinh(gL)\right]^2\right), 
\label{eq:s14}
\end{equation}	
\begin{equation}
P_{i}(L) =P_{s}(0)\left[\frac{\gamma P_{p}}g\sinh(gL)\right]^{2},
\label{eq:s15}
\end{equation}	
where $L$ is the propagating length of the topological waveguide along x
axis. The parametric gain coefficient is given by 
\begin{equation}
g^2=\left[(\gamma P_p)^2-(\kappa/2)^2\right]=-\Delta k\left[\frac{\Delta k}{4} +\gamma P_p\right].
\label{eq:s16}
\end{equation}	
Furthermore, the single gain can be given by~\cite{s5}
\begin{equation}
G_{\mathrm{s}}=\frac{P_{\mathrm{s}}^\mathrm{out}}{P_{\mathrm{s}}^\mathrm{in}}=1+\left(\frac{\gamma P_{\mathrm{p}}}{g}\sinh(gL)\right)^2.
\label{eq:s17}
\end{equation}

\subsection{Right branch: entangled biphoton state generated from intra-band OPP}
After tracing out of inter-band OPP from the FWM process in our diamond-shaped structure (left branch), the FWM process propagating along the right branch can be considered as an entangled biphoton state generator.
Here we start with the nonlinear Hamiltonian in Eq.\ref{eq:s7}, by replacing neglecting the weak terms and pump term $\frac12(\hat{a}_p^{\dagger}\hat{a}_p^{\dagger}\hat{a}_p\hat{a}_p)$, the Hamiltonian of the FWM process can be rewritten by
\begin{equation}
\hat H_{NL}\approx -\hbar \gamma[2(\hat{a}_p^{\dagger}\hat{a}_s^{\dagger}\hat{a}_p\hat{a}_s+\hat{a}_p^{\dagger}\hat{a}_i^{\dagger}\hat{a}_p\hat{a}_i)+(\hat{a}_s^{\dagger}\hat{a}_i^{\dagger}\hat{a}_p\hat{a}_p+\hat{a}_p^{\dagger}\hat{a}_p^{\dagger}\hat{a}_s\hat{a}_i)].
\label{eq:s18}
\end{equation}
Here we apply the electric field to replace the operators as
\begin{equation}
\hat H_{NL}\approx -\hbar \gamma[2(\hat{E}_p^{+}\hat{E}_p^{-}\hat{E}_s^{+}\hat{E}_s^{-}+\hat{E}_p^{+}\hat{E}_p^{-}\hat{E}_i^{+}\hat{E}_i^{-})+(\hat{E}_s^{+}\hat{E}_i^{+}\hat{E}_p^{-}\hat{E}_p^{-}+\hat{E}_s^{-}\hat{E}_i^{-}\hat{E}_p^{+}\hat{E}_p^{+})],
\label{eq:s19}
\end{equation}
in which the pump field operator is considered as the classical field
\begin{equation}
\hat E_p^{(+)}(x,t)=\hat E_p^{-*}(x,t)=A_pe^{i(k_px-\omega_pt)},
\label{eq:s20}
\end{equation}
and the quantized field of signal and idler modes are
\begin{equation}
\hat{E}_{j}^{(-)}(x,t)=\int d\omega_{j}A_{j}^{*}e^{-i(k_{j}x-\omega_{j}t)}\hat{a}_{j}^{\dagger}(\omega_{j}),\quad j=s,i,
\label{eq:s21}
\end{equation}
where the amplitude of the field is $A_j=\sqrt{\frac{\omega_{j}}{4\pi\varepsilon_0n_{j}cA_{eff}}}$. 
By substituting Eq.\ref{eq:s20} and Eq.\ref{eq:s21} into Eq.\ref{eq:s19}, we can obtain the Hamiltonian as 
\begin{equation}
\hat H_{NL}=-\hbar\eta\int_{-\infty}^{\infty}d\omega_s\int_{-\infty}^{\infty}d\omega_i e^{-i(2k_p-k_s-k_i)x}e^{(2\omega_p-\omega_s-\omega_i)t}\hat{a}_{s}(\omega_{s})\hat{a}_{i}(\omega_{i})+h.c.,
\label{eq:s22}
\end{equation}
where the constant term is
\begin{equation}
\eta=\frac{{A_P}^2\gamma}{4\pi\epsilon_0cA_{eff}}\sqrt{\frac{\omega_s\omega_i}{n_sn_i}}.
\label{eq:s23}
\end{equation}
we can calculate the biphoton state generated from the FWM process via first-order perturbation theory by $|\Psi\rangle=\frac{1}{i\hbar}\int_{-\infty}^{\infty}dtH_{NL}|0\rangle$, therefore the biphoton state is given by
\begin{equation}
\begin{aligned}
|\Psi\rangle= \eta
\int\int d\omega_{s}d\omega_{i}\alpha(\frac{\omega_{s}+\omega_{i}}{2})\operatorname{sinc}({\frac{\Delta kL}{2}})\hat{a}_s^{\dagger}{}{(\omega_{s})}\hat{a}_i^{\dagger}{}{(\omega_{i})}|0\rangle,
\end{aligned}.
\label{eq:s24}
\end{equation}
in which the spectrum $\alpha(\frac{\omega_{s}+\omega_{i}}{2})=2\pi\delta(\omega_{s}+\omega_{i}-2\omega_{p})$, and the joint spectral amplitude (JSA) of biphoton state is~\cite{s8} 
\begin{equation}
{\mathcal A}\left(\omega_{s},\omega_{i}\right)=\alpha(\frac{\omega_{s}+\omega_{i}}{2})\operatorname{sinc}({\frac{\Delta kL}{2}}).
\label{eq:s25}
\end{equation}

We use Schmidt decomposition to confirm the entanglement of photon pairs generated via intra-band OPP.
The JSA can be decomposed by\cite{s8}:

\begin{equation}
\mathcal{A}(\omega_{s},\omega_{i})=\sum_{n=1}^{N}\sqrt{\lambda_{n}}\psi_{n}(\omega_{s})\phi_{n}(\omega_{i}),
\label{eq:s26}
\end{equation}
where $\lambda_n$ ($N\in\mathbb{N}$)  represents the Schmidt coefficient,  $\psi_{n}$ and $\phi_{n}$ are are orthonormal functions of $\omega_s$ and $\omega_i$ in the Hilbert space. $\lambda_n$, $\psi_{n}$ and $\phi_{n}$ are connected by these equations

\begin{equation}
\begin{aligned}
\int K_1(\omega,\omega')\psi_n(\omega')d\omega'=\lambda_n\psi_n(\omega),\\
\int K_2(\omega,\omega')\phi_n(\omega')d\omega'=\lambda_n\phi_n(\omega),
\label{eq:s27}
\end{aligned}
\end{equation}
where $K_1$ and $K_2$ are the one-photon spectral correlations, and $\psi_{n}$ and $\phi_{n}$ are corresponding eigenfunctions. When the Schmidt number $N>1$, the biphoton state is considered frequency entangled. The equations can be rewritten as
\begin{equation}
\begin{aligned}
K_{1}(\omega,\omega^{\prime})=\int\mathcal{A}(\omega,\omega_{i})\mathcal{A}^{*}(\omega^{\prime},\omega_{i}){d}\omega_{i},\\
K_{2}(\omega,\omega^{\prime})=\int{\cal A}(\omega_{s},\omega){\cal A}^{*}(\omega_{s},\omega^{\prime}){d}\omega_{s},
\end{aligned}
\label{eq:s28}
\end{equation}
$K_{1}$ and $K_{2}$ form $s\times s$ and $i\times i$ matrices respectively.
The eigenfunctions can be represented as:
\begin{equation}
\begin{aligned}
K_1\psi_n=\lambda_n\psi_n,\\
K_{2}\phi_{n}=\lambda_{n}\phi_{n},
\end{aligned}
\label{eq:s29}
\end{equation}

Eq.\ref{eq:s26} can be rewritten as
\begin{equation}
{\cal A}=\sum_{n=1}^{N}\sqrt{\lambda_{n}}\psi_{n}\phi_{n}^{T},
\label{eq:s29}
\end{equation}

Using Eq.\ref{eq:s29}, the Schmidt coefficients  $\lambda_{n}$ are determined by solving the eigenvalue equations. Notably, frequency entanglement in biphoton states is confirmed when there is more than one non-zero Schmidt coefficient $\lambda_{n}$, or when the entanglement entropy $S_{k}>0$~\cite{s8}. 
Additionally, the entropy of entanglement $S_k$ and Schmidt number $K$  are useful metrics to quantify the degree of entanglement~\cite{s8}
\begin{equation}
S_k=-\sum\limits_{n=1}^N\lambda_n\log_2\lambda_n,
\label{eq:s30}
\end{equation}
\begin{equation}
K=-\frac{\left(\sum\limits_{n=1}^N\lambda_n\right)^2}{\sum\limits_{n=1}^N\lambda_n^2}.
\label{eq:s31}
\end{equation}

The high values of $K=16.24$ and $S_{k}=4.42$ suggest a high quality of high-dimensional frequency entanglement.

\section{Implementing multiple OPPs in honeycomb lattice}
\subsection{Topological honeycomb lattice}
We then consider conducting multiple OPPs in honeycomb lattices that emulate the QVH effect.
We study the valley kink states in topological honeycomb lattices, for undisturbed unit cells with $C_6$ lattice symmetry, degenerate Dirac points appear in the $\rm K$ and $\rm K'$ valleys. The effective Hamiltonian near the $\rm K$ ($\rm K'$) point is expressed as~\cite{s9, s10, s11}
\begin{equation}
H_{K/K^{\prime}}=\tau_z\nu_D(\sigma_x\delta k_x+\sigma_y\delta k_y).
\label{eq:S7}
\end{equation}
Here, $v_{D}$ represents the group velocity, and $\sigma_{x}$ and $\sigma_{y}$ are the Pauli matrices. $\delta\vec{k}=\vec{k}-\vec{k}_{K/K'}$ indicates the deviation of the wavevector. Introducing unit cell distortion ($d_1 \neq d_2$), the Hamiltonian can be modified as follows
\begin{equation}
H_{K/K^{\prime}}=\tau_z\nu_D(\sigma_x\delta k_x+\sigma_y\delta k_y)+\tau_z\gamma\sigma_z.
\label{eq:S8}
\end{equation}
In this expression, $\tau_z=1(-1)$ denotes the $\rm K$ ($\rm K'$) valley pseudospin, $\sigma_{x,y,z}$ denotes the Pauli matrices, $\nu_{D}$ is the group velocity, and $\gamma$ is the strength of the symmetry-breaking perturbation. The perturbations $\gamma{1}$ and $\gamma_2$ are defined as $\gamma_{1}\propto\left[\int_{B}\varepsilon_{z}ds-\int_{A}\varepsilon_{z}ds\right]$ ($\rm VPC1$) and $\gamma_2\propto\left[\int_D\boldsymbol{\varepsilon}_zds-\int_C\boldsymbol{\varepsilon}_zds\right]$ ($\rm VPC2$), respectively, where $\int\varepsilon_zds$ is the integration of the dielectric constant $\varepsilon_z$ at the positions of $A$ and $B$, respectively.
For the given parameters, $d_{A}=0.36a$ and $d_{B}=0.24a$, resulting in $\int_B\varepsilon_zds<\int_A\varepsilon_zds$. Moreover, we find $|\gamma_1|>|\gamma_2|$.

This implies that the modes at the $\rm K$ and $\rm K'$ valleys exhibit opposite circular polarizations: left-handed circular polarization (LCP) and right-handed circular polarization (RCP), respectively. The valley Chern numbers of VPCs are determined by~\cite{s10,s11}:
\begin{equation}
C_{K/K^{\prime}}=\frac{1}{2\pi}\int_{HBZ}\Omega_{K/K^{\prime}}(\delta\vec{k})dS=\pm1/2,
\label{eq:S9}
\end{equation}
where $\Omega=\nabla_k\times\vec{A}(k)$ is the Berry curvature, and $\vec{A}(k)$ is the Berry connection. This integration region covers half of the Brillouin zone. Thus, the disparity in the valley Chern numbers of the system is calculated as $|C_{K/K'}|=1$, confirming the topological characteristics of VPCs.
These findings indicate that the oscillation patterns at the $\rm K$ and $\rm K'$ valleys show different polarizations. Specifically, the K valley exhibits a left-handed circular polarization (LCP) while the K' valley shows a right-handed circular polarization (RCP).

\begin{figure*}
\centering
\includegraphics[width=1\textwidth]{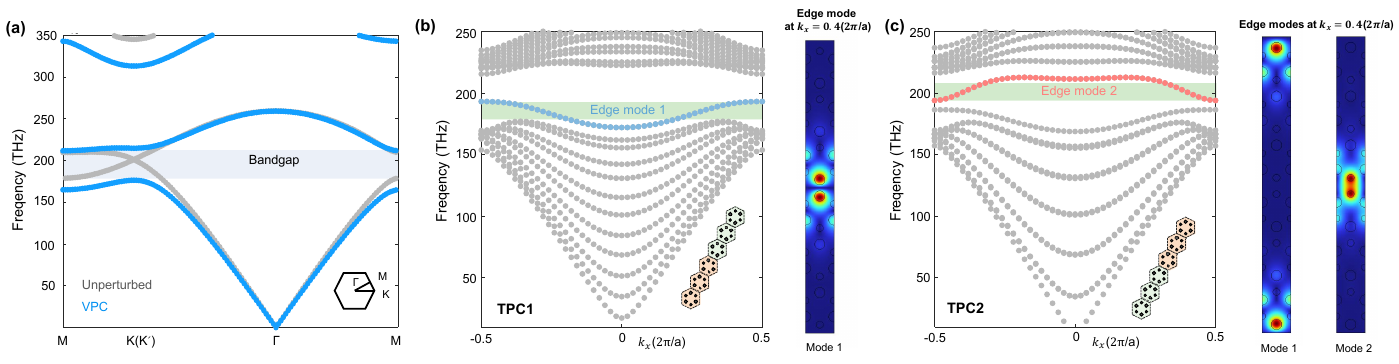}
\caption{(a) Band diagrams for an unperturbed honeycomb lattice (grey dots) and an expanded one (blue dots). 
Dispersion relations for (b) TPC1 and (c) TPC2 respectively, which are composed of shrunken ($d=0.24a$) and expanded ($d=0.36a$) honeycomb lattices. 
The right insets show the electric field distributions for different edge modes.}
\label{fig:S2}
\end{figure*}

\subsection{Multiple OPPs in VPCs}

In VPCs, topologically protected edge states, also known as valley kink states~\cite{s5}, are observable at the boundary between VPC1 and VPC2.
As depicted in Fig.\ref{fig:S2}(a), we observe two opposite valley kink states, which are linked to a different valley. 
Only one mode exists for a specific wavevector within the bandgap, and it is localized at the inner interface.
Exchanging the two VPCs leads to the inversion of valley Chern numbers and further facilitates the inversion of topological edge states.
Correspondingly, for the reversed VPC, the electric fields of these edge modes are localized differently: one at the outer interface and the other at the inner interface (Fig.\ref{fig:S2}(b)-(c)).
Similarly, mode matching simplifies the coupling of edge modes from the sandwich VPC to other VPCs, offering convenience for the implementation of multiple OPPs.

\begin{figure}[h]
\centering
\includegraphics[width=1\textwidth]{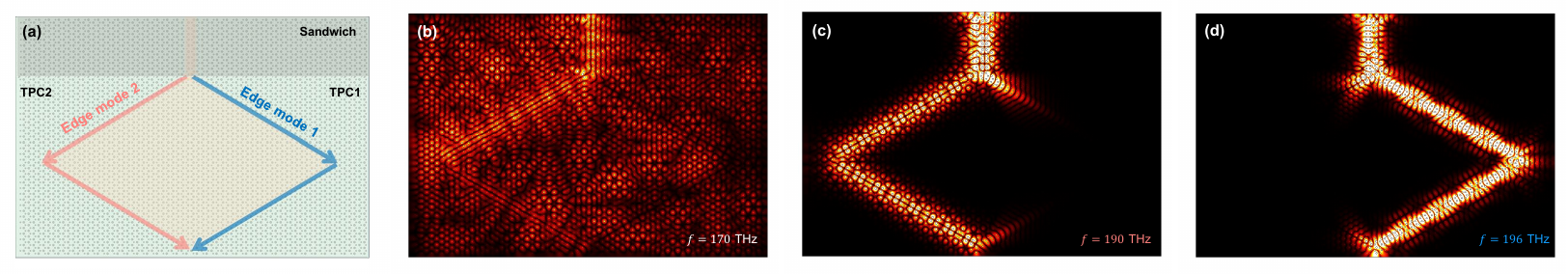}
\caption{(a) A design of a topological apparatus featuring a sandwich waveguide (depicted in grey) and a diamond-shaped setup. 
(b)-(d) Field profiles for edge modes at different frequencies in our topological device.}
\label{fig:s3}
\end{figure}

In our study, we demonstrate a resembling diamond-shaped structure composed of both trivial (green area) and nontrivial (orange area) VPCs, as illustrated in Fig.\ref{fig:s3}(a). 
Due to the mirror symmetry of VPCs, the two edges of the rhombus correspond to two topological edge states for distinct VPCs. 
We simulate the field profiles of field distributions in this diamond-shaped structure at different frequencies. 
As shown in Fig.\ref{fig:s3}(b)-(d), these edge modes can be efficiently transmitted to their respective ports within the different frequency spectrum.
This design also results in frequency division functionality, which can be applied to separating quantum states.

We use a Gaussian pump with a central frequency of  $f_p=190$ THz and a full width at half-maximum (FWHM) of $\Delta f_p=115 $ GHz.
The JSA is depicted in Fig.7(e) in the main text, where the main intensity along the anti-diagonal axis indicates a signal-idler frequency correlation, and two additional bright spots correspond to inter-band OPP interaction between two edge modes.

\subsection{OPA and entangled biphoton generation from different OPPs}	
\begin{figure}[h]
\centering
\includegraphics[width=0.6\textwidth]{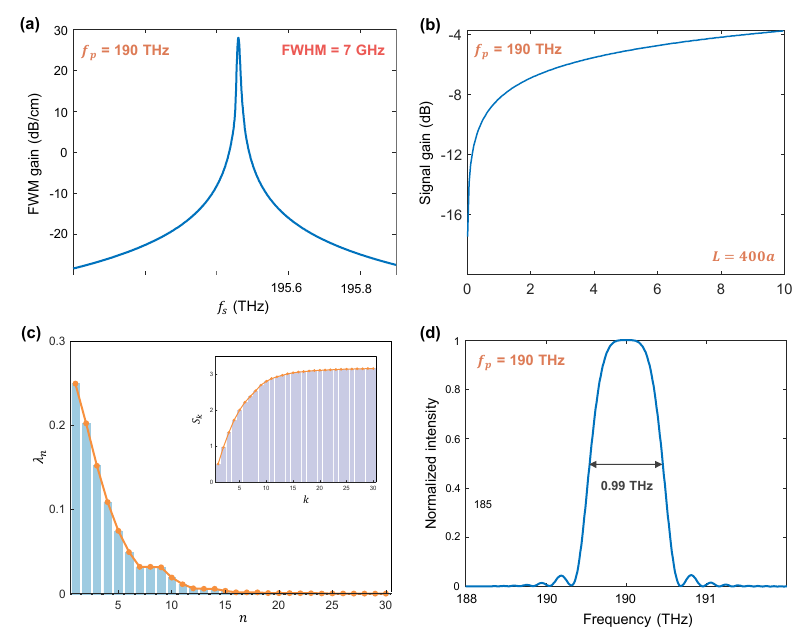}
\caption{(a) FWM gain coefficient for intra-band OPP at the 190 THz pump frequency for a $400a$ length topological waveguide. 
(b) Signal gain as a function of the pump power.
(c) Normalized Schmidt coefficients $\lambda_n$ andentanglement entropy $S_k$ for the biphoton state generated form inter-band OPP.
(d) Normalized two-photon spectral distribution at the 190 THz pump frequency.}
\label{fig:s4}
\end{figure}

In this section, we show that our topological honeycomb lattice can be utilized to implement two functionalities: OPA and the generation of entangled photon pairs.
First, we consider the OPA through FWM with inter-band OPP.
The frequency division in the diamond-shaped structure results in the spatial separation of signal photons, facilitating the straightforward extraction of amplified optical signals. 
We first examine the frequency distribution of signal and idler modes resulting from inter-band OPP under varying pump frequencies.
Figure \ref{fig:s4}(a) shows the FWM gain coefficient for intra-band OPP at the 190 THz pump frequency in a $400a$ length topological sandwich waveguide (1 W pump power). 
Intra-band OPP excites a super-narrow bandwidth of significant amplification, with a full width at half maximum (FWHM) of about 7 GHz.
At the central frequency of the amplification region, a FWM gain coefficient of up to 30 dB/cm is attainable.
Such OPA with a tunable narrow bandwidth is particularly useful for amplifying signals from a single photon source.
In Fig. \ref{fig:s4}(b), the signal gain is illustrated as a function of the pump power.

In addition, we conduct the generation and control of a frequency-entangled biphoton state originating from the inter-band OPP. 
Notably, the pump, signal, and idler modes can all be coupled into the left boundary of the diamond-shaped TPC structure. 
This setup is advantageous for directly extracting broadband entangled photon pairs from this boundary.
We apply Schmidt decomposition to assess the separability of the JSA~\cite{s8}.
Figure \ref{fig:s4}(c) displays the distributions of normalized Schmidt coefficients $\lambda_n$ and entanglement entropy $S_k$, respectively. 
For our topological quantum state, the theoretical values calculated for the Schmidt number and entanglement entropy are $K=6.67$ and $S_{k}=3.16$, respectively, demonstrating the presence of a high-quality frequency-entangled biphoton state in our sandwich VPCs.
We then calculate the normalized two-photon spectral distribution at the pump frequency of 190 THz. 
Figure \ref{fig:s4}(d) illustrates that the 3dB bandwidth of the two-photon spectrum is 0.99 THz.
These features show a key feature of our high-dimensional topological quantum entangled state.

\begin{figure*}
\centering
\includegraphics[width=1\textwidth]{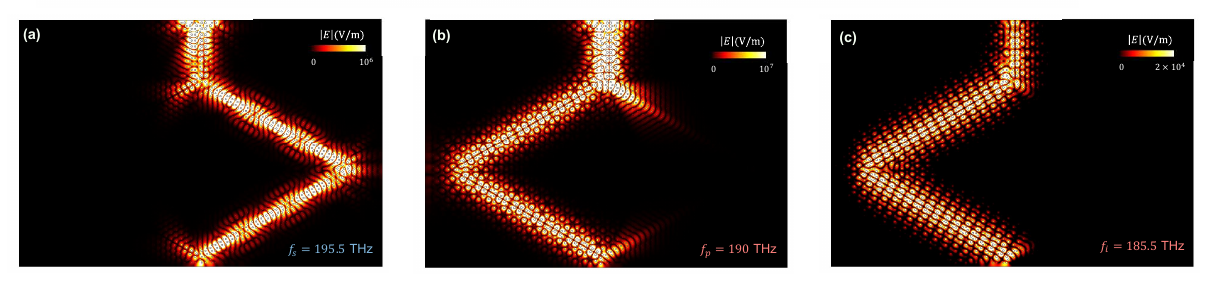}
\caption{Field profiles of the FWM process in the diamond-shaped VPC structure at the frequencies of the (a) signal mode ($f_s=195.5$ THz), (b) pump mode ($f_p=190$ THz), and (c) idler mode($f_i=184.5$ THz), respectively. }
\label{fig:s5}
\end{figure*}

Furthermore, we conduct simulations of the FWM in a diamond-shaped VPC structure with CW pump excitation.
The frequencies for the pump, signal, and idler modes are set at  $f_s=195.5$ THz, $f_p=190$ THz, and $f_i=184.5$ THz, respectively.
As illustrated in Fig.\ref{fig:s5}(a)-(c), the field profiles at the idler frequency demonstrate the FWM process within the topological edge modes.
Notably, due to their different frequencies, the signal mode is coupled to the left side of the diamond-shaped structure, while the pump and generated idler mode are directed to the right side.

%